\documentclass[reprint,
 amsmath,amssymb,
 aps,
 prb,
]{revtex4-1}

\usepackage{graphicx}
\usepackage{dcolumn}
\usepackage{bm}

\begin{document}

\title{Superspreading events suggest aerosol transmission of SARS-CoV-2 by accumulation in enclosed spaces}

\author{John M. Kolinski}
\email{john.kolinski@epfl.ch}
\author{Tobias M. Schneider}%
 \email{tobias.schneider@epfl.ch}
\affiliation{%
 Institute of Mechanical Engineering, \'{E}cole Polytechnique F\'{e}d\'{e}rale de Lausanne,
Lausanne, 1015, Switzerland
}%

\date{\today}

\begin{abstract}
Viral transmission pathways have profound implications for public safety; it is thus imperative to establish a complete understanding of viable infectious avenues. Mounting evidence suggests SARS-CoV-2 can be transmitted via the air; however, this has not yet been demonstrated. Here we quantitatively analyze virion accumulation by accounting for aerosolized virion emission and destabilization. Reported superspreading events analyzed within this framework point towards aerosol mediated transmission of SARS-CoV-2. Virion exposure calculated for these events is found to trace out a single value, suggesting a universal minimum infective dose (MID) via aerosol that is comparable to the MIDs measured for other respiratory viruses; thus, the consistent infectious exposure levels and their commensurability to known aerosol-MIDs establishes the plausibility of aerosol transmission of SARS-CoV-2. Using filtration at a rate exceeding the destabilization rate of aerosolized SARS-CoV-2 can reduce exposure below this infective dose.
\end{abstract}

\maketitle


\section{Introduction}

The infectious pathways of a virus determine its course through a host population. Whether a virus is transmitted is fundamentally governed by the virus shedding rate of an infected host, the minimal infective dose (MID) required to infect, and the transmission pathway mediating virus material delivery.  Despite clear guidelines on social distancing and lock-down policies intended to disrupt transmission of SARS-CoV-2 by respiratory droplet transmission\cite{noauthor_advice_nodate}, SARS-CoV-2 has proven to be a challenging virus to successfully contain due in part to its prolonged symtomless incubation period\cite{chu_comparative_2020,furukawa_evidence_2020}. Ongoing developments in the guidance regarding possible aerosol transmission reflect uncertainty in the role of aerosols in viral transmission; indeed, mask-wearing is marginally effective in reducing viral transmission\cite{zhang_identifying_2020}, but not all infectious particles are captured by typical surgical masks\cite{kahler_fundamental_2020}, complicating a direct assessment of the role of aerosols in transmitting SARS-CoV-2. While there is some evidence for the aerosol transmission of the virus\cite{morawska_airborne_2020,prather_reducing_2020}, the circumstances under which aerosol transmission might be expected are currently unknown.

Here we quantitatively analyze several reported superspreading events\cite{hamner_high_2020,jang_early_nodate,pung_investigation_2020,lu_covid-19_2020,gunther_investigation_nodate,yusef_early_nodate}. In each of these 20 recent superspreading events, the primary source of the infection would not likely have been in near- or intimate contact with the susceptible population, thus mitigating the likelihood of infection via the respiratory droplet transmission pathway that is addressed by social distancing guidelines\cite{noauthor_advice_nodate}.
Estimating room volume and co-occupancy time for all events, allows us to to use a simple mass-conservation based model for aerosol-born virus to calculate the exposure levels, and test whether the quantitative data suggests aerosol-mediated transmission of SARS-CoV-2.  

\begin{figure}[!h]
\includegraphics[width=0.8\linewidth]{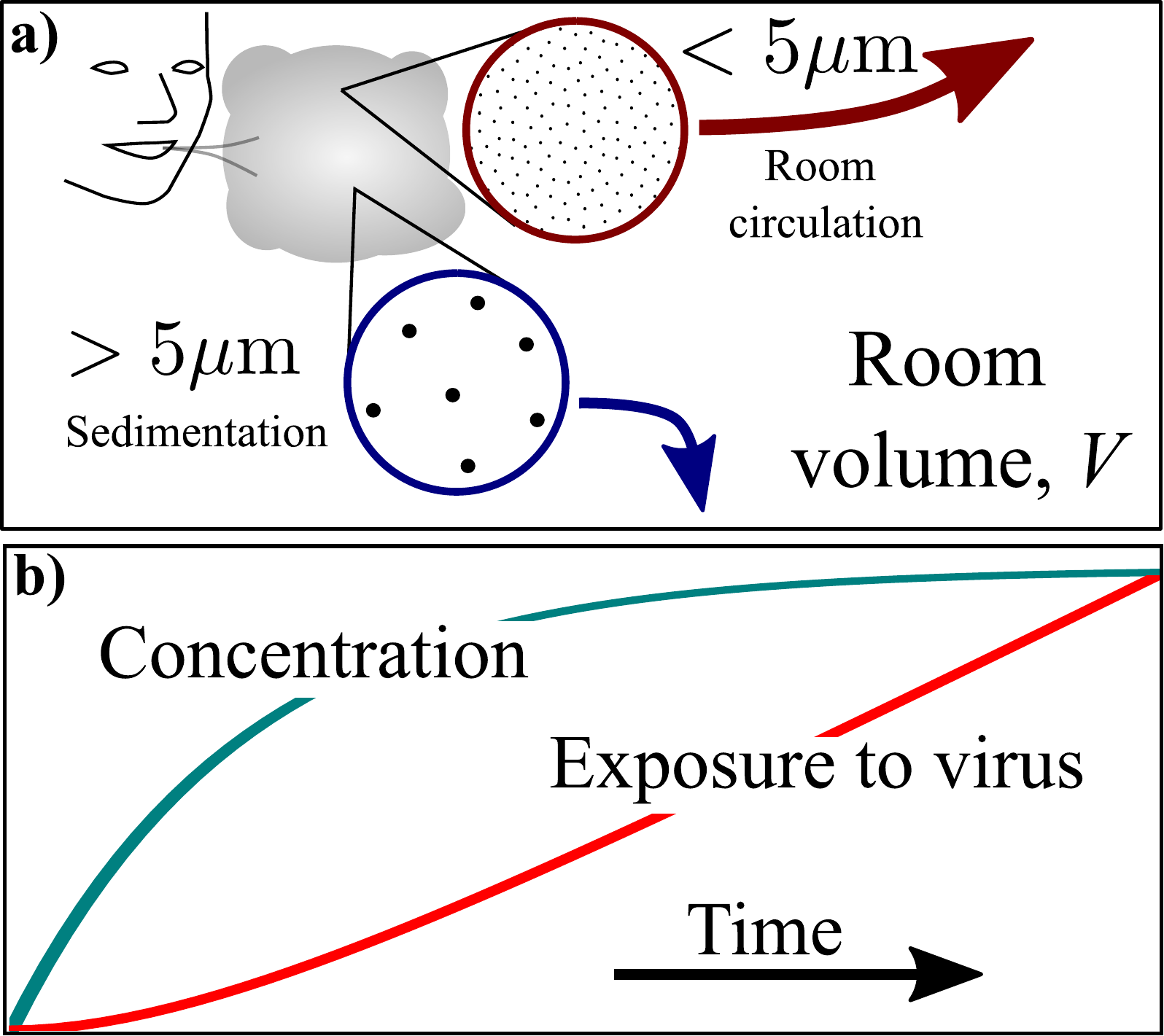}
\caption{Aerosolization of viral cargo in an enclosed space.
(a) A schematic of a closed room with volume $V$. An infected person can emit \emph{both} rapidly sedimenting respiratory droplets and aerosol-sized particles when breathing, speaking, coughing or sneezing. Particles with a diameter of 5 $\mu$m sediment in quiescent air at a rate of 3 m/hour. These particles are readily dispersed by room currents set up by air conditioning, thermal gradients and background flow of the room air. (b) For aerosolized virus emitted into an enclosed space, the rate of emission will ultimately be balanced by the destabilization rate; these dynamics lead to the evolution of aerosolized virus concentration plotted here. A non-infected occupant breathing at a constant rate in the same space is exposed to $N_{exp}$ particles over time (right axis).}
\label{fig1}
\end{figure}

\section{Results}

The pathway for transmission of respiratory viruses via droplets depends on their size, as depicted schematically in Fig.~\ref{fig1} a). Small droplets with radius $a$ below 2.5 microns settle in quiescent air at a velocity set by the balance of the gravitational force $F_g = 4/3\, \pi \,\rho \,g \,a^3$ and the Stokes drag force, $F_d = 6 \pi a \mu v$. Such a droplet descends at a velocity $v$ of about 2 \textit{m/hour}; however, prevailing currents in a typical room driven by convection, ventilation or movement of room occupants significantly exceed this settling velocity. Thus, aerosolized droplets will remain suspended for hours, and disperse widely in an occupied room within minutes\cite{morawska_airborne_2020,prather_reducing_2020}. Due to this rapid and thorough mixing of aerosols, for the aerosol transmission pathway, the physical distance between persons in an enclosed space becomes irrelevant; exposure to the airborn virus is controlled only by the concentration of virus in the air. This is in contrast to transmission of the virus by large respiratory droplets, as these droplets fall to the ground in seconds, limiting the spatial range of infectivity\cite{bourouiba_turbulent_2020,mittal_flow_2020,somsen_small_2020}. 

Motivated by these considerations of droplet suspension velocity, we assume that the flow within the room under consideration is well-mixed within minutes; this is fast compared to the aerosolized SARS-CoV-2 destabilization rate $\gamma$. The measured values for the decay of viral titer of SARS-CoV-2 suggest that the aerosolized viral titer decays by a factor of 1/e in one hour\cite{van_doremalen_aerosol_2020}. We use 5 microns as a threshold for the diameter of an aerosol droplet, corresponding to the calculated sedimentation rate of 2 \textit{m/hour}; larger droplets are assumed to sediment prior to infecting via an airborn pathway. Because the air is well mixed, the complex spatio-temporal evolution of aerosol distribution can be represented by the uniform virion concentration $C(t)$, which is a function of time only, as has been assumed in long-standing prior analyses for airborn disease transmission\cite{riley1961airborne}. 

To study the effect of virus accumulation in closed and unfiltered environments, we use straightforward conservation laws of aerosol-born virions to formulate an expression for the time-rate-of-change of the volumetric concentration of aerosolized virions $C(t)$ at time $t$ in terms of a viral source $s$, in number of aerosol-born viral particles shed per time, and the destabilization rate $\gamma$, as $ \frac{d C(t)}{dt} = s/V - \gamma C(t)$. Here $V$ is room volume. The expression $C(t) = \frac{s}{\gamma V} (1-e^{-\gamma t})$ describes the temporal evolution of $C$, as shown in Fig.~\ref{fig1} b). If the room air is filtered or replenished, a third term $\gamma_{filt} C(t)$ is subtracted from the right-hand side of this equation yielding an effective destabilization rate that combines the natural decay and filtration. 

\begin{figure}[!h]
\includegraphics[width=\linewidth]{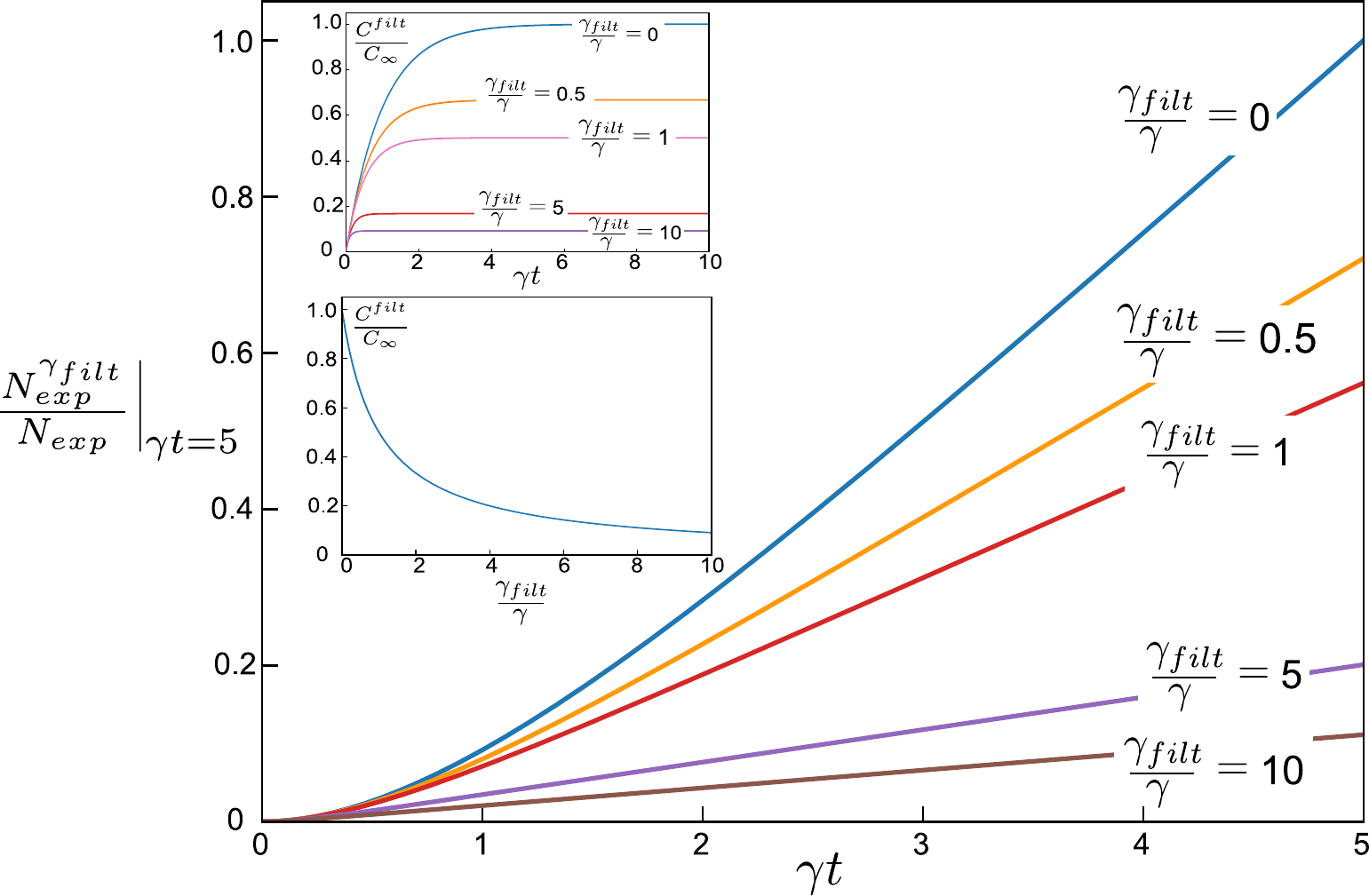}
\caption{Analysis of air filtration for virion accumulation. (inset, top) Air filtration modifies the equation presented in the main text such that $\partial C / \partial t = s / V - \gamma C -\gamma_{filt} C$. We define a composite rate $\tilde{\gamma} = \gamma + \gamma_{filt}$, which can be directly substituted into our original solution. Here the temporal dynamics of the concentration in a room with filtration $C^{filt} (t)$ normalized by the unfiltered steady-state concentration $C_{\infty}$ is plotted for several filtration rates. Time is normalized by the destabilization rate of the virus. The background concentration is significantly reduced by filtration rates typical in hospitals, which can exceed 10 volumes / hour\cite{cheng_air_nodate}. (inset, bottom) The normalized steady-state concentration of the virus with filtration $C_{\infty}^{filt} / C_{\infty}$ is plotted as a function of $\gamma_{filt}$. Filtration can significantly reduce the background virus, but the effect is not pronounced until the filtration rate is comparable to or larger than the destabilization rate of the virus. (main figure) Filtration clearly reduces the rate of virion exposure for room occupants, as shown here by $N_{exp}$, curves plotted as a function of normalized occupancy time for a given room volume. }
\label{figfilt}
\end{figure}

Filtration acts to reduce the aerosolized SARS-CoV-2 concentration. We explore the role of filtration by adding an additional decay term to the rate of concentration change with a rate $\gamma_{filt}$, measured in room volumes per time. This directly reduces the steady-state concentration of virions in a given room, as shown in Fig.~\ref{figfilt} (inset, top). Notably, the filtration rate has to exceed the intrinsic decay rate of the virus to make a pronounced difference on the steady-state concentration, as shown in Fig.~\ref{figfilt} (inset, bottom). As a consequence of filtration at a rate greater than the intrinsic decay rate of the virus, the overall virion exposure can be significantly reduced, as the exposure is integrated over the time of occupancy; this effect is clearly visible in Fig.~\ref{figfilt} (main figure).

Is an occupant likely to be infected at such concentrations? To answer this question, we calculate the virion exposure $N_{exp}$, recognizing that it must exceed a minimum infective dose (MID), which remains unknown for SARS-CoV-2. $N_{exp}$ is given by the time-integral of the respiration rate of a room occupant, $\dot{Q}$, as $N_{exp} = \int_0^{T} C(t) \dot{Q} d t$, where $\dot{Q}$ is the respiration rate. The average adult at rest takes 12 breaths per minute, cycling a volume of 0.5 liters / breath; this amounts to a total respiration rate of 360 liters per hour, or $\dot{Q} = 0.36 m^3$ / hour. Thus, $N_{exp}$ with a single infected occupant in the above cases is $14$ and $108$ per hour, respectively. We see the pronounced effect of the accumulation timescale when we evaluate $N_{exp}$ for 30 minutes: it is 4 and 31 particles, respectively; thus it is less than 1/3 the exposure for twice that time period. 

\begin{figure}[!h]
\includegraphics[width=\linewidth]{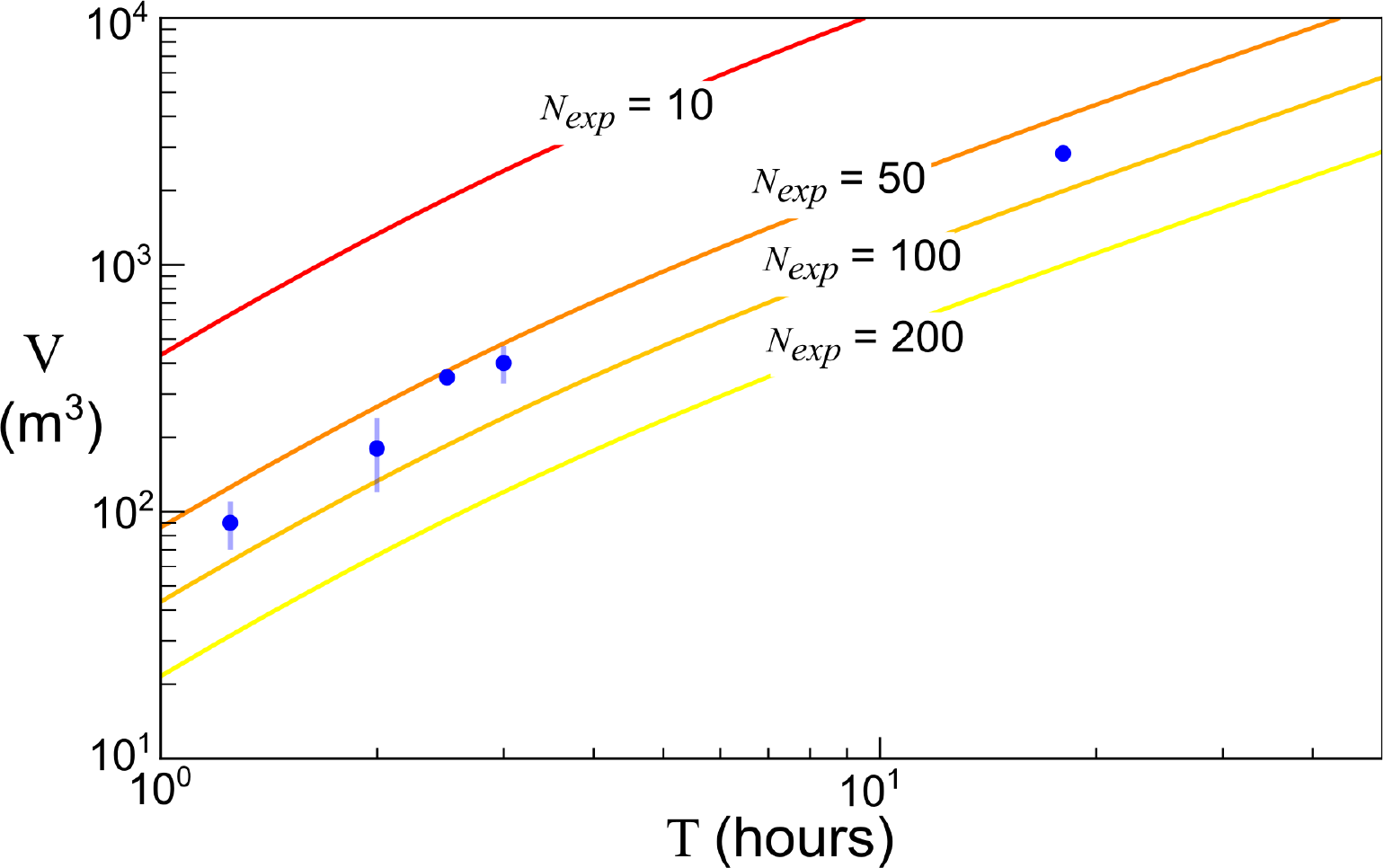}
\caption{$N_{exp}$, the number of viral particles a room occupant is exposed to, is calculated as a function of room volume $V$ and occupancy time $T$. $N_{exp}$ is indicated on each curve. Data from several super-spreading events\cite{hamner_high_2020, pung_investigation_2020, lu_covid-19_2020, gunther_investigation_nodate} with a single spreading source in non-exercise scenarios are plotted on the graph, with error bars indicated for the events. Details of these events, and assumptions used in the calculation of the iso-$N_{exp}$ curves are included in Methods.}
\label{fig2}
\end{figure}

This analysis can be generalized for arbitrary times and room volumes; we present several iso-$N_{exp}$ curves as a function of time and room volume in Fig.~\ref{fig2}. While The minimum infectious dose (MID) for aerosolized SARS-CoV-2 is currently unknown, at room volumes that are typically encountered in daily activities, and over timescales of co-occupancy of the order of hours, we observe iso-$N_{exp}$ values that fall within the range of the MIDs reported for other aerosol-transmissible viruses including the aggressive Influenza A (H2N2), adenoviruses and SARS-CoV-1\cite{yezli_minimum_2011,schroder_covid-19_2020,watanabe_development_2010}, pointing to the possibility of aerosolized transmission of SARS-CoV-2.  

To asses the viablity of aerosolized SARS-CoV-2 transmission despite its unknown MID, we analyse reported super-spreading events that comply with our analytical framework, and either directly report or offer a means of reasonably estimating the room volume $V$ and cooccupancy time $T$, as described in appendix~\ref{appa} on the determination of parameters used in our model\cite{hamner_high_2020,jang_early_nodate,pung_investigation_2020,lu_covid-19_2020,gunther_investigation_nodate,yusef_early_nodate}.

The $V$ and $T$ values for the sub-set of events that involve a single viral shedder at a resting respiratory rate are directly compared with the calculated iso-$N_{exp}$ values plotted in Fig.~\ref{fig2}. Among these five events, the data fall within a remarkably consistent range of $N_{exp}$ values between 50 and 100, as shown in Fig.~\ref{fig2}. The tight range of $N_{exp}$ values emerging from analysis of super-spreading events suggests a possible unique MID via aerosol transmission for SARS-CoV-2 of $N_{exp} \approx 50$.


\begin{figure}[!h]
\includegraphics[width=\linewidth]{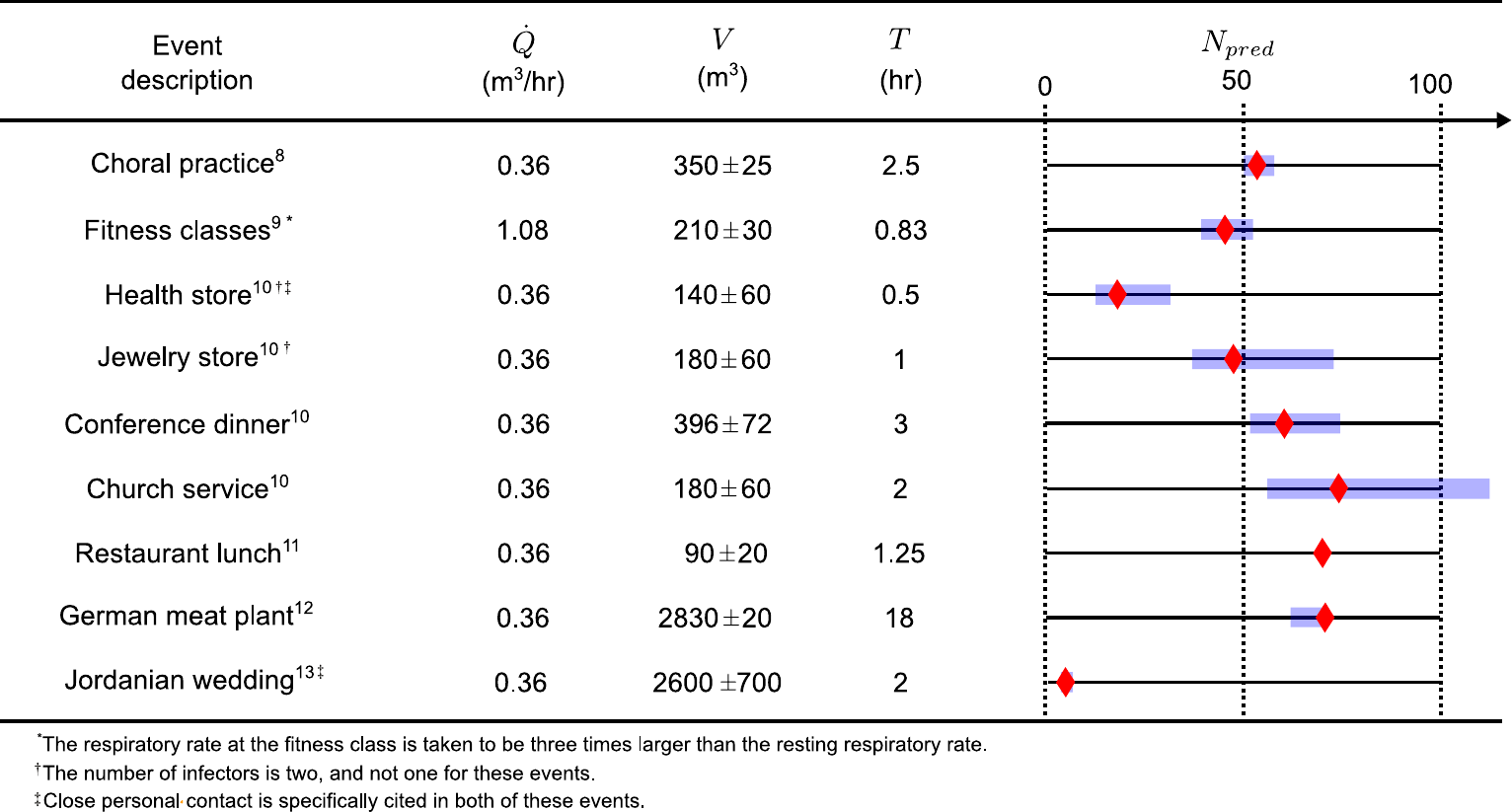}
\caption{Tabulated data for several super-spreading events. A total of 20 distinct superspreading events\cite{hamner_high_2020,jang_early_nodate,pung_investigation_2020,lu_covid-19_2020,gunther_investigation_nodate,yusef_early_nodate} for SARS-CoV-2 are analyzed, and the parameter values used to formulate the prediction for numerical value of viral particle exposure $N_{exp}$ are presented. The predicted value for $N_{exp}$ is plotted on the graph at the right. The predicted value is shown using a red diamond, and the range of predicted values are shown using the blue rectangles. A detailed discussion of how the range is established is provided in Methods. Note that there were two sources for the events at the health products shop and the jewelry shop\cite{pung_investigation_2020}, and the estimated respiratory rate is elevated for the fitness classes\cite{jang_early_nodate}; these directly alter the $N_{pred}$.}
\label{fig3}
\end{figure}

If the value of $N_{exp} \approx 50$ is indeed the MID for aerosol transmission of SARS-CoV-2, it should independently arise under further analysis. We test this hypothesis by extending our analysis to other super-spreading events with identical co-occupancy with different respiration rates\cite{jang_early_nodate} or number of spreaders\cite{pung_investigation_2020}. Using the same analysis framework, but accounting for modifications of the source number or respiration rate as described in appendices~\ref{appa}~\&~\ref{appb}, we tabulate and graph the predictions of $N_{exp}$, including the data from Figure 2; altogether, this analysis includes over 200 infections and over 1000 exposed persons. We find that with the exception of two events that involve documented close physical contact\cite{pung_investigation_2020,yusef_early_nodate}, all cases fall within the anticipated range of $N_{exp} \approx 50$ virions, as can be seen in Fig.~\ref{fig3}, underscoring the importance of a unique critical exposure threshold for aerosol transmission. 

\section{Discussion and Limitations} 

Our model incorporates two key parameters: the first is the rate of aerosolized virion shedding by an infected person $s$, inferred from data for other coronaviruses\cite{leung_respiratory_2020} and second, the virion destabilization rate $\gamma$, measured for SARS-CoV-2\cite{van_doremalen_aerosol_2020}. Each of the parameters used in the model were determined as described in detail in appendix~\ref{appb}. We additionally assume that filtration is slow compared to the viral destabilization rate. However, this framework can account the importance of filtration if the filtration rate is known, as we present in detail in Fig. S1. If the volumetric exchange of air is greater than one or two room volumes per hour, $C_{\infty}$ can be reduced by a factor of two. Filtration that captures aerosol particles is atypical in most commonly encountered environments; indeed, only specialized filters will capture aerosol particles, such as HEPA filters, or ventilation with air outside of the building. In hospital environments, where the air filtration rate can exceed 10 volumes per hour, our model suggests that $N_{exp}$ over several hours can be reduced approximately 10-fold. Such a reduction would potentially reduce $N_{exp}$ below the MID for SARS-CoV-2. This might explain why exposed medical workers in a hospital environment were not infected\cite{faridi_field_2020,cheng_air_nodate}.

The least well-constrained parameters in this analysis are the shedding rate of aerosol-born virions and the inferred value of the infective dose in aerosol form. However, these two properties are properties of the virus, and should thus be independent of the specific circumstances of each superspreading event. Indeed, our analysis suggests that these properties are universal for SARS-CoV-2, and points towards the same infective exposure level in all events. 
Using the typical shedding value of SARS-CoV-1 within the analysis of aerosol transmission by accumulation, superspreading events suggest minimum infective doses of SARS-CoV-2 commensurate with other infectious viruses, including the influenza-A (H2N2) strand that caused the 1957-'58 influenza pandemic\cite{yezli_minimum_2011,schroder_covid-19_2020}.
Indeed, our study suggests that in terms the infective dose for aerosolized transmission, SARS-CoV-2 behaves much like a particular influenza - unfortunately, the H2N2 influenza strand that generated a global pandemic. The evidence for the potential airborn transmission for SARS-CoV-2 is by now sufficient to form a scientific consensus\cite{morawska_airborne_2020,prather_reducing_2020,fennelly_particle_2020}; however, the significance of this transmission mode has not yet been directly demonstrated.

In the course of preparing this manuscript, we became aware of a similar approach taken by another research group that reaches similar conclusions with a more granular treatment of the emission rate estimates - the value for the parameter $s$ in our work - that are determined for SARS-CoV-2 emission in aerosol form from recent studies\cite{Prentiss}; these measurements confirm that the range of $s$ based on SARS-CoV-1 measurements\cite{leung_respiratory_2020} is comparable to that of SARS-CoV-2.

Under the condition that virions are transported by aerosol droplets, data from several reported superspreading events all indicate the same narrow range of infectious dose. This points towards the practical relevance of aerosol transmission of SARS-CoV-2. 

\appendix
\section{Obtaining values for $V$,  $T$ and the number of shedders for superspreading events \label{appa}}
The formulation of our model requires an accurate determination of the room volume $V$, the co-occupancy period $T$ and the number of aerosol-based virion shedders to calculate $N_{exp}$. While some studies report these values, for example by including full spatial maps of the environment\cite{gunther_investigation_nodate}, others report partial data for $V$ without the room height\cite{jang_early_nodate, lu_covid-19_2020}, or no data at all for $V$ other than a venue description\cite{hamner_high_2020, pung_investigation_2020, yusef_early_nodate}. Typically, values for $T$ were very well constrained, including co-occupancy times down to the minute, recorded with CCTV footage of the incident in one case\cite{lu_covid-19_2020}. In total, the data collected comprise 20 discrete events, with all but the fitness class comprising a single event. The fitness classes involved 12 different events, driven by the simultaneous infection of several fitness class instructors at a training event prior to the several classes held in the following week\cite{jang_early_nodate}. The data evaluated in this manuscript correspond to a total of 207 infections with over 1100 documented persons exposed.

There was typically a single documented infector for the superspreading events analyzed here. For the tour group, there were likely two infectors from a co-travelling sub-group of a 20 person tour (cluster A)\cite{pung_investigation_2020}. A single infector is most likely for all the other events, although the morning shift workers at the German meat processing facility were exposed by a primary infector who worked and lived with a potential secondary infector (labeled B1 and B2), where the first co-exposure took place on the day prior to the three days of total exposure during shift work\cite{gunther_investigation_nodate}. The Jordanian wedding event was triggered by a primary source who had close contact with a large number of participants in two days prior to the two hour wedding, and likely had close contact with a large number of attendees during the wedding itself, as the primary source was the bride’s father\cite{yusef_early_nodate}; as such, non-aerosol pathways cannot be ruled out and are indeed likely to have contributed to the scale of the outbreak. A short duration encounter with two sources in Singapore also included close inter-personal contact at a health products shop, where interactions likely took place via person to person contact (cluster A, health products shop)\cite{pung_investigation_2020}; thus, also in this event non-aerosol mediated infection is likely.

Values for V were reported directly\cite{gunther_investigation_nodate}, calculated from the reported floor area\cite{jang_early_nodate, lu_covid-19_2020}, with an estimated ceiling height range, or estimated / calculated from alternative sources\cite{hamner_high_2020, pung_investigation_2020, yusef_early_nodate}. For the German meat processing facility the full dimensions of the shared shift space were provided; we used only the proximal floor space in our analysis, as walls separated this volume from the remainder of the plant\cite{gunther_investigation_nodate}. The Jordanian wedding took place at an indoor facility designed to accommodate approximately 400 guests\cite{yusef_early_nodate}; estimating approximately 16 m$^2$ per 10 persons yields a floor area of approximately 640 m$^2$; with a room height of 3 to 5 meters, we obtain a volume estimate of approximately 2600 $\pm$ 700 m$^3$. Similar considerations were used to determine the size of the conference dinner facility in cluster B for a total of 100 conference attendees\cite{pung_investigation_2020}. The choral practice event took place in Skagit county, Washington\cite{hamner_high_2020}. The choir’s website provides photos of the choral practice room prior to the outbreak, enabling measurements of the floor area with an uncertainty of a few meters per dimension\cite{skagitweb}. Neither the jewelry store, the health products shop, nor the church included values of V or floor space\cite{pung_investigation_2020}; in each of these cases, between 5 and 10 google street view images of stores belonging to these categories were analyzed, and appropriately wide error bounds were used for V. The Korean fitness class events reported a room size of approximately 60 m$^2$\cite{jang_early_nodate}, and a floorplan was provided for the Chinese restaurant lunch\cite{lu_covid-19_2020}. In both cases, room heights were not provided and are estimated between 3 and 4 meters. For the Chinese restaurant lunch, the total restaurant area was served by two primary air conditioners. We selected the zone of the restaurant served by the air conditioner where the infected person sat, as this was a distinct zone within the restaurant\cite{lu_covid-19_2020}.

The period of co-occupancy $T$ was reported in either minutes or fractions of an hour, and is thus tightly constrained for each event. For the German meat processing plant, three morning shifts over a span of three days were reported as exposed. A total of two hours of breaks were reported during these shifts; thus we use a 6 hour period per shift\cite{gunther_investigation_nodate}. The total period of exposure thus constitutes 18 hours. In our analysis, we explicitly ignore any effects outside of these shifts including metabolism or decay of virions during off hours, for example.

The number of superspreading events we analyze is not comprehensive, as there are significantly more superspreading events that have been identified in the literature; however, it is not common practice to include the precise size of the space, or even the time duration of the interaction, and thus these events are excluded from this analysis. Furthermore, a significantly number of better documented superspreading events were clearly associated with close inter-personal contact (e.g. families, roommates, etc.) and thus are not compatible with the analytical framework presented here.

\section{Determination of parameter values. \label{appb}} 

There are 3 primary parameters that are used to calculate $N_{exp}$: $s$, $\dot{Q}$ and $\gamma$. $s$ represents the source strength in units of aerosolized virions per hour, $\dot{Q}$ is the respiratory volume per hour and $\gamma$ is the proportion of aerosolized virions that destabilize in one hour.

$s$ is currently unknown for SARS-CoV-2. In order to constrain this parameter, we evaluate the number of virions emitted per hour on average for three other human corona viruses: HCoV-NL63, HCoV-OC43 and HCoV-HKU1\cite{leung_respiratory_2020}. Droplets with a diameter of less than 5 microns were collected using an apparatus called the Gesundheit II during normal respiration of infected individuals over a 30 minute window. According to the virus reported in the subject’s nasal swab, the sample collected was analyzed using RT-PCR specific to this virus, offering a direct, quantitative measurement of virus copies\cite{leung_respiratory_2020}. Within the sampling of these viruses, all three were represented in aerosol shedding, and a total of 4 of 10 total infected persons shed aerosolized virus in this study\cite{leung_respiratory_2020}. The number of virions shed varied from 660 to 57,000 in 1/2 hour; 16,300 were shed on average, corresponding to an average value of s = 32,600 virions/hour, as extracted from Fig. 1 (a) in Leung\cite{leung_respiratory_2020}. This value of $s$ is used throughout the manuscript. Notably, $s$ might vary depending on the activity reported - for example, singing\cite{hamner_high_2020} or exercise\cite{jang_early_nodate} might augment the value of $s$; as we have no way of determining how s could vary under these circumstances, we retain the same value in all calculations. It is furthermore interesting to note that both the aerosol virus shedding rates, and the fraction of infected persons shedding aerosolized virus, are commensurate for three analyzed corona viruses and two infuenza viruses\cite{leung_respiratory_2020}, suggesting that there could be similarities between these families of viruses, as was previously noted\cite{morawska_airborne_2020}.

The respiratory rate $\dot{Q}$ for an adult person is known to be 12 breaths per hour, and the volume per breath is 0.5 \textit{l} on average. This corresponds to $\dot{Q}$ = 0.36 m$^3$/hour; this value is used for all calculations except the high-intensity fitness class\cite{jang_early_nodate}, where an elevated $\dot{Q} = 1.08$ m$^3$/hour is used, corresponding to a three-fold higher respiratory rate. Notably, the accepted minute respiratory rate of 6 \textit{l/min} is less than values typically used for adults doing light work or activity, which are larger, at 10 \textit{l/min}\cite{riley1962infectiousness, catanzaro1982nosocomial, nardell1991airborne, fennelly1998relative}. This would lead to proportionally larger exposure levels than those calculated here.

$\gamma$ was extracted directly from data collected for aerosolized SARS-CoV-2\cite{van_doremalen_aerosol_2020}. In a one hour period, the viral titer drops by approximately $e^{-1}$, setting $\gamma=1$/hour. A second study suggested that aerosolized virus can remain infectious for a far longer period\cite{fears_early_nodate}; however, these data do not convey a convincing trend, and the data are less dense than in the study used here to determine $\gamma$\cite{van_doremalen_aerosol_2020}.

%

\end{document}